\documentclass[reprint, prx, superscriptaddress, floatfix]{revtex4-1}

\usepackage{savesym}
\usepackage{hyperref}
\usepackage{graphicx}
\usepackage{amsmath}
\savesymbol{comment}
\usepackage{wasysym}
\usepackage{amsfonts}
\usepackage{color}
\usepackage{url}

\usepackage{verbatim}
\usepackage{upgreek}
\usepackage{bm}
\usepackage{units}
\usepackage{placeins}
\usepackage{babel}
\restoresymbol{TXF}{comment}
\usepackage{float}
\usepackage[draft]{changes}

\newcommand{\beq}{\begin{equation}}
\newcommand{\eeq}{\end{equation}}
\newcommand{\bea}{\begin{eqnarray}}
\newcommand{\eea}{\end{eqnarray}}
\newcommand{\bal}{\begin{align}}
\newcommand{\eal}{\end{align}}

\begin{document}
\title{Lattice dynamics and ultrafast energy flow between electrons, spins, and phonons \newline in a 3{\sl d} ferromagnet}
\author{Daniela Zahn} 
 \email{zahn@fhi-berlin.mpg.de}
\affiliation{Fritz-Haber-Institut der Max-Planck-Gesellschaft, Faradayweg 4-6, 14195 Berlin, Germany}
\author{Florian Jakobs}
\affiliation{Freie Universit\"at Berlin, Arnimallee 14, 14195 Berlin, Germany}

\author{Yoav William Windsor}
\affiliation{Fritz-Haber-Institut der Max-Planck-Gesellschaft, Faradayweg 4-6, 14195 Berlin, Germany}
\author{H\'{e}l\`{e}ne Seiler}
\affiliation{Fritz-Haber-Institut der Max-Planck-Gesellschaft, Faradayweg 4-6, 14195 Berlin, Germany}
\author{\mbox{Thomas Vasileiadis}}
\altaffiliation[Present address: ]{Faculty of Physics, Adam Mickiewicz University, Wieniawskiego 1, 61-712 Pozna\'{n}, Poland}
\affiliation{Fritz-Haber-Institut der Max-Planck-Gesellschaft, Faradayweg 4-6, 14195 Berlin, Germany}
\author{\mbox{Tim A. Butcher}}
\altaffiliation[Present address: ]{School of Physics, Trinity College, Dublin 2, Ireland}
\affiliation{Helmholtz-Zentrum Dresden-Rossendorf, Bautzner Landstra\ss{}e 400, 01328 Dresden, Germany}
\author{Yingpeng Qi}
\altaffiliation[Present address: ]{Center for Ultrafast Science and Technology, School of Physics and Astronomy, Shanghai Jiao Tong
University, 200240 Shanghai, China}
\affiliation{Fritz-Haber-Institut der Max-Planck-Gesellschaft, Faradayweg 4-6, 14195 Berlin, Germany}

\author{Dieter Engel}
\affiliation{Max-Born-Institut, Max-Born-Stra\ss{}e 2A, 12489 Berlin, Germany}
\author{Unai Atxitia}
\affiliation{Freie Universit\"at Berlin, Arnimallee 14, 14195 Berlin, Germany}
\author{Jan Vorberger}
\affiliation{Helmholtz-Zentrum Dresden-Rossendorf, Bautzner Landstra\ss{}e 400, 01328 Dresden, Germany}
\author{Ralph Ernstorfer}
 \email{ernstorfer@fhi-berlin.mpg.de}
\affiliation{Fritz-Haber-Institut der Max-Planck-Gesellschaft, Faradayweg 4-6, 14195 Berlin, Germany}

\begin{abstract}

The ultrafast dynamics of magnetic order in a ferromagnet are governed by the interplay between electronic, magnetic and lattice degrees of freedom. In order to obtain a microscopic understanding of ultrafast demagnetization, information on the response of all three subsystems is required. 
A consistent description of demagnetization and microscopic energy flow, however, is still missing. Here, we combine a femtosecond electron diffraction study of the ultrafast lattice response of nickel to laser excitation with ab initio calculations of the electron-phonon interaction and energy-conserving atomistic spin dynamics simulations. 
Our model is in agreement with the observed lattice dynamics and previously reported electron and magnetization dynamics. Our approach reveals that the spin system is the dominating heat sink in the initial few hundreds of femtoseconds and implies a transient non-thermal state of the spins.
Our results provide a clear picture of the microscopic energy flow between electronic, magnetic and lattice degrees of freedom on ultrafast timescales and constitute a foundation for theoretical descriptions of demagnetization that are consistent with the dynamics of all three subsystems.

\end{abstract}

\maketitle
\date{\today}

\section{Introduction}

The discovery of ultrafast demagnetization in ferromagnetic nickel in 1996 by Beaurepaire et al. \cite{1996Beaurepaire} induced a paradigm shift in the field of magnetism. The experiment proved that magnetic order can be manipulated on femtosecond time scales, therefore offering new perspectives in data storage. Since then, researchers have worked towards a microscopic understanding of the phenomenon \cite{1996Beaurepaire,2010Kirilyuk,2010Koopmans,2010Battiato,Toews2015,2014Mueller,Kazantseva2007,2008Balashov,2009Krauss,2010Schmidt,2010Atxitia,2015Carpene,2017Eich,2019Chen}. To acquire microscopic insights into the processes governing the ultrafast demagnetization in itinerant 3{\sl d} ferromagnets, knowledge about the response of electronic, magnetic and lattice degrees of freedom to laser excitation is required. 
Most of the experimental work in literature focuses either on the magnetization dynamics using the time-resolved magneto-optical Kerr effect (tr-MOKE) \cite{1996Beaurepaire,2006Cinchetti,2007DallaLonga,2008Carpene,2010Koopmans,2012Roth,2012Vorakiat,2013Schellekens,2016Turgut,2018You} or time-resolved X-ray magnetic circular dichroism (tr-XMCD) \cite{2007Stamm,2009Kachel,2013Eschenlohr}, or on the electronic response using time-resolved photoemission methods \cite{2006Cinchetti,2017Eich,2018Tengdin,2018Gort}. In contrast, the lattice response has received only limited attention \cite{Dornes2019,2019Maldonado,2019Ritzmann,2008Wang,2010Wang}. Knowledge of the lattice dynamics is essential, as it plays several important roles in the dynamics of the system: First of all, it serves as a sink for angular momentum \cite{Dornes2019}. Second, in addition to receiving angular momentum, the lattice is also an energy sink: it drains energy from the electronic system on ultrafast timescales via the creation of phonons. Hence, the electron-phonon coupling strength strongly influences the energy content of the electronic system and consequently also the magnetization dynamics. Finally, the lattice response is in turn also influenced by the magnetization dynamics, both during the demagnetization and the magnetization recovery (remagnetization). The demagnetization of an isolated sample requires spin excitations, e.g. spin flips and/or magnons, which cost energy. This is also visible in the equilibrium heat capacity, which shows a divergence at the Curie temperature \cite{1981Meschter}. Due to this energy cost, ultrafast demagnetization reduces the energy content in the electronic system and thus indirectly influences the lattice dynamics as well.

Several models have been developed and used to describe the magnetization dynamics of 3{\sl d} ferromagnets following laser excitation \cite{1996Beaurepaire,2010Koopmans,2010Battiato,2014Mueller,Kazantseva2007,Ma2012,Manchon2012}. 
In addition to the magnetization dynamics, however, a consistent model should also describe the electronic and lattice responses correctly. In particular, due to the relatively large heat capacity of the lattice, an accurate description of electron-lattice equilibration is important. Nonetheless, literature values for the electron-phonon coupling parameter $G_\mathrm{ep}$ of nickel vary by more than an order of magnitude \cite{1998Wellersdorf,1996Beaurepaire,Saito2003,Caffrey2005,Hopkins2007,2008Lin,2010Koopmans,2013Rethfeld,2014Dvornik,2016Persson,2018Tengdin,2019Ritzmann}. So far, experimental studies of ultrafast lattice heating in nickel have mostly employed optical techniques \cite{2005vanKampen,Caffrey2005,Hopkins2007}, which are sensitive to both the electronic and the lattice responses. The most direct technique to study the lattice is diffraction, but there are only few studies that measured the lattice heating directly with time-resolved diffraction \cite{2008Wang,2010Wang}. In addition, the electron-phonon coupling was often deduced from observables without considering the energy cost of demagnetization \cite{1998Wellersdorf,Caffrey2005,Hopkins2007,2010Koopmans,2016Persson}. The large spread in literature values for $G_\mathrm{ep}$ can manifest itself in an imprecise description of the electron-lattice equilibration and makes different models less comparable.

To obtain a consistent model for the microscopic energy flow and the magnetization dynamics, it is paramount to compare theoretical results to the response of all three subsystems, including the lattice. At the same time, the energy flow dynamics between the subsystems need to be described consistently. In particular, energy flow to and from magnetic degrees of freedom needs to be considered. Regarding the existing demagnetization models, the microscopic three-temperature model (M3TM) introduced by \mbox{Koopmans et al. \cite{2010Koopmans}} as well as conventional micromagnetic and atomistic spin dynamics simulations \cite{Atxitia2007,2012Chimata,Evans2015,2010Atxitia} disregard the energy flow associated to the magnetization dynamics. In contrast, the three-temperature model (3TM) introduced by Beaurepaire et al. takes energy flow to and from the spin system into account \cite{1996Beaurepaire,2014Kimling}. 
However, to deduce the three different coupling constants of the 3TM reliably from experimental data, information on the response of more than one subsystem is required. In addition, the 3TM describes the spin system based on its properties in thermal equilibrium, which is a questionable assumption on short time scales after laser excitation \cite{Kazantseva2007,2018Tengdin}.
Similarly, a modified version of the M3TM includes energy flow to and from the spin system, but also assumes a thermalized spin system \cite{2012Roth}. Dvornik et al.~introduced an energy-conserving model that goes beyond a thermal description of the spin system by employing micromagnetic simulations \cite{2014Dvornik}, but no direct comparison with experimentally measured lattice dynamics has been made yet.

In this work, we fill this gap by providing a comprehensive experimental and theoretical description of the lattice dynamics in ferromagnetic nickel. We use femtosecond electron diffraction (FED) to directly measure the lattice response to laser excitation. In Section II we provide an overview of the electron diffraction experiment and the experimental results. The excellent time resolution of our electron diffraction setup allows us to resolve the lattice heating in nickel on femtosecond time scales. Section III discusses the comparison between the experimental results and energy flow models of increasing complexity. For this comparison, we perform spin-resolved density functional theory (DFT) calculations to obtain the electron-phonon coupling parameter $G_\mathrm{ep}$ as well as the electronic and lattice heat capacities. In Section III.A, we compare the experimental results to the commonly used two-temperature model (TTM) and a modified TTM with strong electron-spin coupling (s-TTM). The latter is the minimal extension of the TTM that considers magnetic degrees of freedom. This comparison reveals that energy transfer to and from magnetic degrees of freedom has a strong impact on the lattice dynamics. In Section III.B, we go a step further and aim for a quantitative description not only of the lattice dynamics, but of all three subsystems using energy-conserving atomistic spin dynamics (ASD) simulations. This hybrid approach of spin dynamics simulations and energy flow model is shown to provide a consistent description of both the non-equilibrium dynamics of the spin system as well as the energy flow between the different subsystems. Section IV provides a summary of the key findings.

\section{Experiment}

The samples were freestanding, polycrystalline nickel films with a thickness of \unit[20]{nm} sandwiched between \unit[5]{nm} layers of Si$_3$N$_4$ on both sides to avoid oxidation. They were prepared on NaCl crystals by magnetron sputter deposition at room temperature. To obtain freestanding samples, the thin films were transferred onto standard TEM grids using the floating technique \cite{2007Dwyer}. The samples were not exposed to a magnetic field before the measurements. 

\begin{figure}[bth]
\begin{center}
\includegraphics[width=\columnwidth]{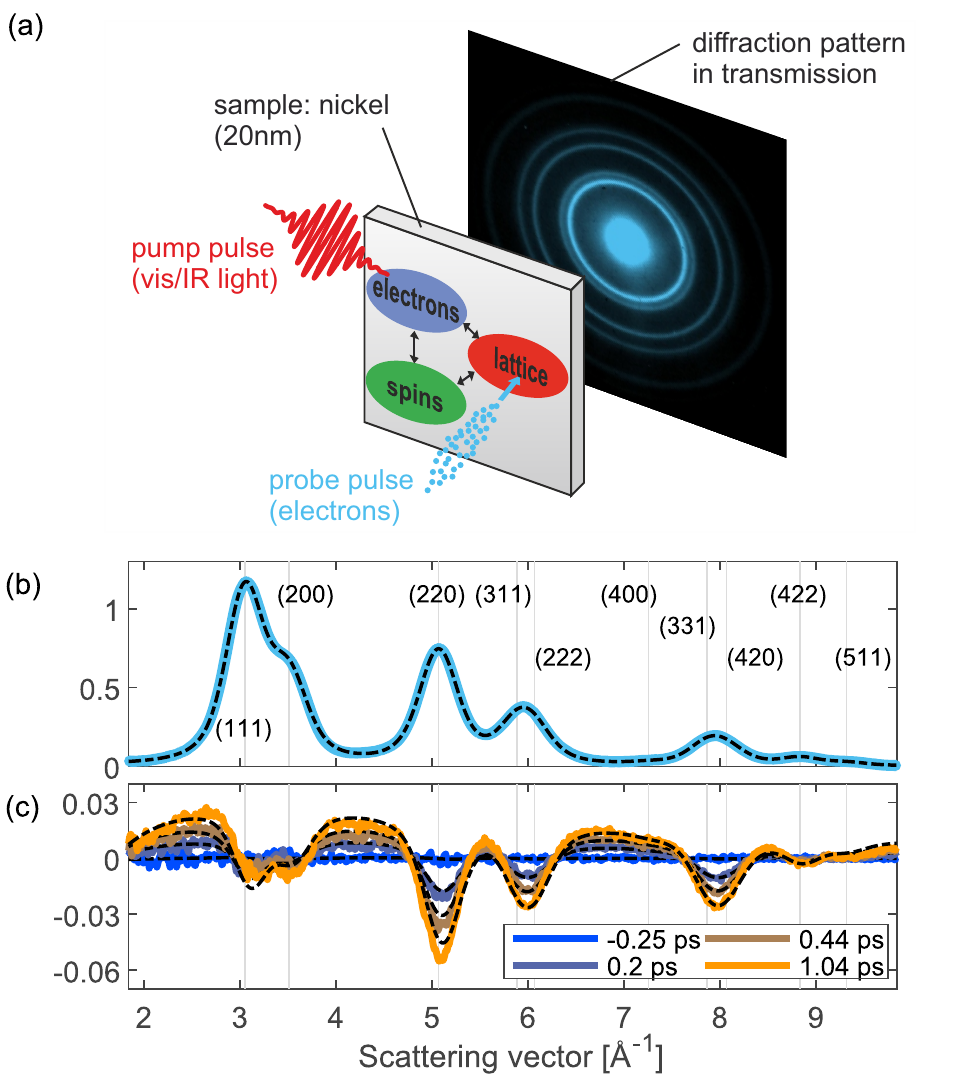}
\caption{Details of the femtosecond electron diffraction experiment. (a) Schematic diagram of the experiment. The electrons in the sample are excited using a visible or infrared laser pulse. The excited electrons transfer energy to the spins as well as to the lattice, depending on the respective coupling strengths (black arrows). The lattice response is probed using an ultrashort electron pulse, which diffracts off the sample. Diffraction patterns are recorded in transmission. b) Radial average of the diffraction pattern (solid blue line) before laser excitation. The dashed black line is a fit to the data (static fit). The background contribution obtained from the static fit was subtracted. c) Differences of the radial averages at several pump-probe delays (solid lines) compared to the radial average before laser excitation. The dashed black line shows the fits to the data (dynamic fit). The details of the fits are described in the text.}
\label{fig:intro}
\end{center}
\end{figure}

To study the ultrafast structural dynamics of nickel, we used the compact femtosecond electron diffractometer described in Ref.~\cite{2015Waldecker}. The samples were excited using ultrashort (ca.~\unit[50-80]{fs} FWHM) laser pulses with different wavelengths (\unit[2300]{nm}, \unit[770]{nm} and \unit[480]{nm}), at \unit[4]{kHz} repetition rate. The measurements were conducted at room temperature (\unit[295]{K}). The structural response of the sample was probed in transmission using short electron pulses. The kinetic energy of the electrons was \unit[65-77]{keV}, depending on the experiment. In total, the temporal resolution achieved in the experiments was around \unit[170]{fs}. Figure~\ref{fig:intro}(a) illustrates the measurement principle and shows a diffraction pattern of our polycrystalline nickel sample.

To analyze the changes in the diffraction pattern after laser excitation, the recorded images were radially averaged. A typical radial average of our nickel samples is displayed in Fig.~\ref{fig:intro}(b) (solid blue line). Next, we performed a fit to the radial averages. Here we apply a global fitting approach \cite{our_Pt}, which extracts the lattice dynamics based on the full diffraction pattern instead of individual Bragg reflections as conventionally done \cite{Ligges2009,Nakamura2016}. In the first step of the fitting routine (static fit), we fitted the average of all radial averages before laser excitation to a function consisting of Lorentzian peaks plus a background function, all convolved with a Gaussian. The peak amplitudes of the Lorentzians were adjustable but the peak positions were fixed in the fit, except for a parameter for the conversion of pixels to scattering vector, a parameter accounting for aberrations of the electron lens and small correction factors for the individual peaks ($\leq$\unit[5]{\%} deviation). The peak width was one fit parameter, i.e. it was the same for all peaks. The fit result is displayed in Fig.~\ref{fig:intro}(b) (dashed black line). We used the range from the Bragg reflections (111) to (511), as shown. From the Bragg reflection intensities, we deduce that the sample has a preferred orientation, but this does not affect our analysis of the lattice dynamics. In the second step of the fitting routine, which yields the lattice dynamics after laser excitation (dynamic fit), we fixed all parameters of the fit function at the values obtained from the static fit, except the change in atomic mean-squared displacement (MSD), the lattice expansion and the background parameters, and fitted all the radial averages of the measurement. The MSD is related to the peak intensities as follows \cite{Peng}:
\begin{equation}
    \frac{I(t)}{I_0}=\mathrm{exp}\{ -\frac{1}{3}\hspace{2pt}q^2\hspace{2pt}\Delta\langle u^2\rangle\ \}
\end{equation}
Here, $q$  is the scattering vector, $\Delta\langle u^2\rangle$ is the MSD change and $I_0$ is the intensity before laser excitation. Figure~\ref{fig:intro}(c) shows changes of the radial averages after laser excitation for several pump-probe delays together with the fit results of the dynamic fit (dashed black lines). The fit yields the evolution of the MSD as a function of pump-probe delay, which is then converted into lattice temperature using the tabulated Debye-Waller factor of Ref. \cite{Peng}. The deviations of the fit results from the experimental data are caused by secondary scattering effects and the limitations of the phenomenological background function. They do not influence the timescales of the extracted lattice dynamics. The precision of the lattice dynamics is determined using the standard error from the fit. The corresponding error bars are shown as grey shaded areas in all figures. Further details about the global fitting routine are described in Ref.~\cite{our_Pt}.

\begin{figure}[bth]
\begin{center}
\includegraphics[width=1.0\columnwidth]{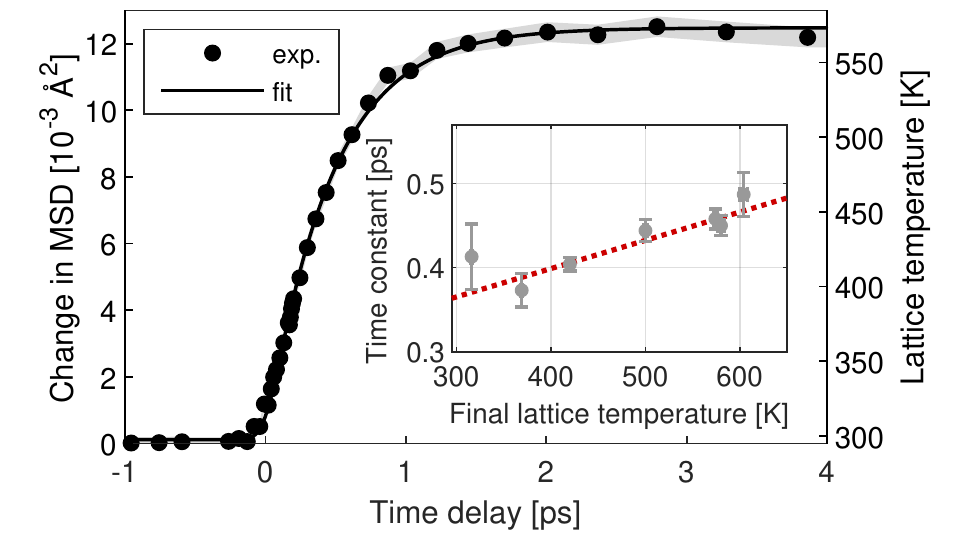}
\caption{Time evolution of the mean squared displacement (MSD) and the lattice temperature after laser excitation with \unit[2300]{nm} light. In this measurement, the absorbed energy density was \unit[1230]{$\frac{\mathrm{J}}{\mathrm{cm}^3}$}. The black dots show the experimental data and the black line is a fit with a single exponential function, convolved with a Gaussian (FWHM: \unit[170]{fs}) to account for the time resolution. The grey shaded area represents the standard errors of the data points, obtained from the fit of the radial averages.  The inset shows the time constants (fit result) for different excitation densities. The error bars represent the standard errors of the single exponential fits. The dotted red line is a linear fit to the data ($\tau=a\cdot(T_\mathrm{final}-\unit[295]{\mathrm{K}})+b$), with $a=0.336\pm\unit[0.06]{\frac{\mathrm{fs}}{\mathrm{K}}}$ and $b=360\pm\unit[20]{\mathrm{fs}}$. The errors of $a$ and $b$ are the standard errors from the fit.}
\label{fig:results2300}
\end{center}
\end{figure}

Figure~\ref{fig:results2300} shows the evolution of the MSD and the lattice temperature as a function of pump-probe delay for a pump wavelength of \unit[2300]{nm} (\unit[0.54]{eV}).
The temperature rise can be well described by a single exponential function, convolved with the instrument response of \mbox{\unit[{\raise.17ex\hbox{$\scriptstyle\sim$}}\hspace{1.5pt}170]{fs}}. The inset of Fig.~\ref{fig:results2300} shows the time constants of the single exponential function (fit results) for different fluences. The time constant is found to increase linearly with excitation density (dotted red line). Our time resolution of around \unit[170]{fs} enables us to resolve the lattice heating. We observe time constants that are significantly faster than previous electron diffraction reports \cite{2010Wang,2008Wang}. The experimental data as well as the results for the MSD and lattice temperature dynamics are available on a data repository \mbox{\cite{zenodo}}. The TTM and s-TTM results discussed in the next section are also available there.

\section{Results and energy flow models}
\subsection{Two-temperature models}

To go beyond a phenomenological description of the lattice dynamics and connect our observations to microscopic quantities, a model is required. For non-magnetic materials, a frequently used model is the TTM \cite{1974Anisimov,1987Allen}, which describes the time evolution of the system by considering the lattice and the electrons as two coupled heat baths. 
\begin{figure}[bth]
\begin{center}
\includegraphics[width=\columnwidth]{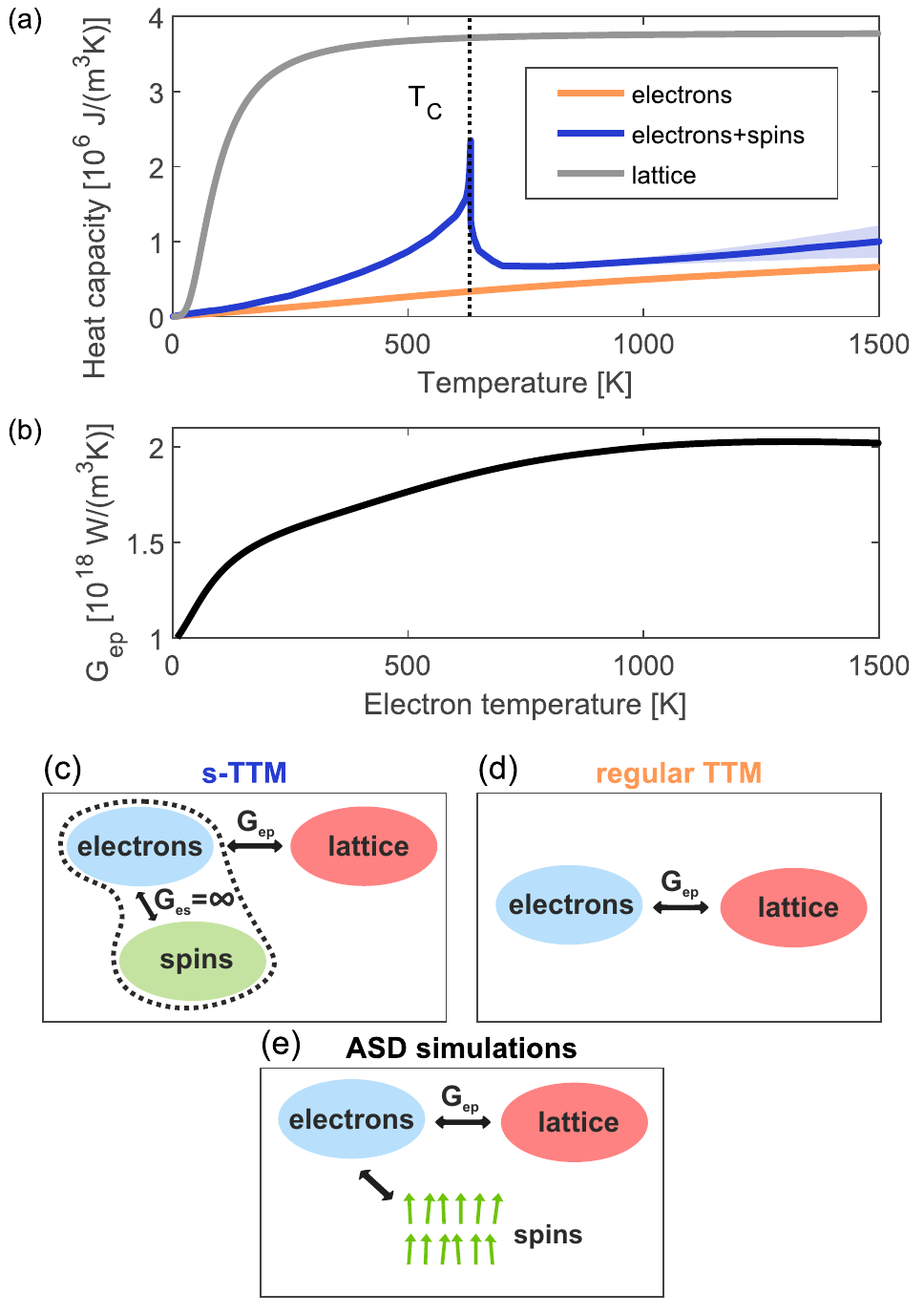}
\caption{Temperature-dependence of model parameters and schematic diagrams of the models. (a) Heat capacities of the electron (orange) and lattice subsystems (grey) as well as the combined heat capacity of electrons and spins (blue). Electronic and lattice heat capacities are calculated based on the spin-resolved DFT results. Since the magnetic contribution to the heat capacity cannot be calculated using DFT, we use the combined heat capacity of electrons and spins determined from experiments \cite{1981Meschter} for the s-TTM. The magnetic contribution peaks at the Curie temperature $T_\mathrm{c}$ (vertical dotted line). The light blue shaded area corresponds to the error estimate. (a) Electron-phonon coupling parameter $G_\mathrm{ep}$ as a function of electron temperature, obtained from the spin-resolved DFT calculations. The sum of majority and minority $G_\mathrm{ep}$ is shown. Panels~(c), (d) and (e) are schematic diagrams of  the s-TTM, the regular TTM and the ASD simulations, respectively (see text for details).}
\label{fig:heat_capacities}
\end{center}
\end{figure}
In magnetic materials, such an approach neglects the magnetic degrees of freedom. However, they have a non-negligible contribution to the total heat capacity, as shown in  Fig.~\ref{fig:heat_capacities}(a). Several approaches have been introduced to take into account energy flow to and from magnetic degrees of freedom \cite{1996Beaurepaire,2010Wang,2014Dvornik}. Here, we follow the approach of Refs. \cite{2010Wang,2018Tengdin} and consider electronic and magnetic degrees of freedom as one heat bath with a common temperature. In the following, we refer to the magnetic contribution as "spins" for simplicity. Note that this includes the orbital magnetic moment. The TTM equations are modified in the following way:
\beq
c_\mathrm{l}(T_\mathrm{l})\cdot\frac{dT_\mathrm{l}}{dt}=G_\mathrm{ep}(T_\mathrm{es})\cdot(T_\mathrm{es}-T_\mathrm{l})
\label{eq:TTM1}
\eeq
\beq
[c_\mathrm{e}(T_\mathrm{es})+c_\mathrm{s}(T_\mathrm{es})]\cdot\frac{dT_\mathrm{es}}{dt}=G_\mathrm{ep}(T_\mathrm{es})\cdot(T_\mathrm{l}-T_\mathrm{es})+S(t),
\label{eq:TTM2}
\eeq
with $G_\mathrm{ep}$: electron-phonon coupling, $T_\mathrm{l}$: lattice temperature, $T_\mathrm{es}$: temperature of electrons and spins, $c_\mathrm{l}$: lattice heat capacity, $c_\mathrm{e}$: electron heat capacity, $c_\mathrm{s}$: spin heat capacity, $S(t)$: source term (laser excitation).

Figure~\ref{fig:heat_capacities}(c) shows a schematic diagram of this modified TTM (s-TTM) and Figure~\ref{fig:heat_capacities}(d) visualizes the regular TTM for comparison. The only difference between the two models is that in the case of the s-TTM, the spin heat capacity is added to the electronic heat capacity. For this we used the combined heat capacity of electrons and spins provided by Ref. \cite{1981Meschter} (blue curve of \mbox{Fig.~\ref{fig:heat_capacities}}(a)). The electron-phonon coupling parameter $G_\mathrm{ep}$, shown in Fig.~\ref{fig:heat_capacities}(b),
as well as the heat capacity of the lattice (grey curve of Fig.~\ref{fig:heat_capacities}(a)), were obtained using spin-resolved DFT calculations. The details of the calculations are described in Appendix~A. For the comparison of the s-TTM to a regular TTM we also calculated the heat capacity of the electrons from the DFT calculations (orange curve of Fig.~\ref{fig:heat_capacities}(a)). To compare the two models to the experimentally measured lattice response, we determined the absorbed energy densities based on the lattice temperature in the range \unit[1.5-4]{ps} and the heat capacities. The arrival time of the laser pulse was determined from the exponential fits described earlier. 

\begin{figure}[bth]
\begin{center}
\includegraphics[width=1.0\columnwidth]{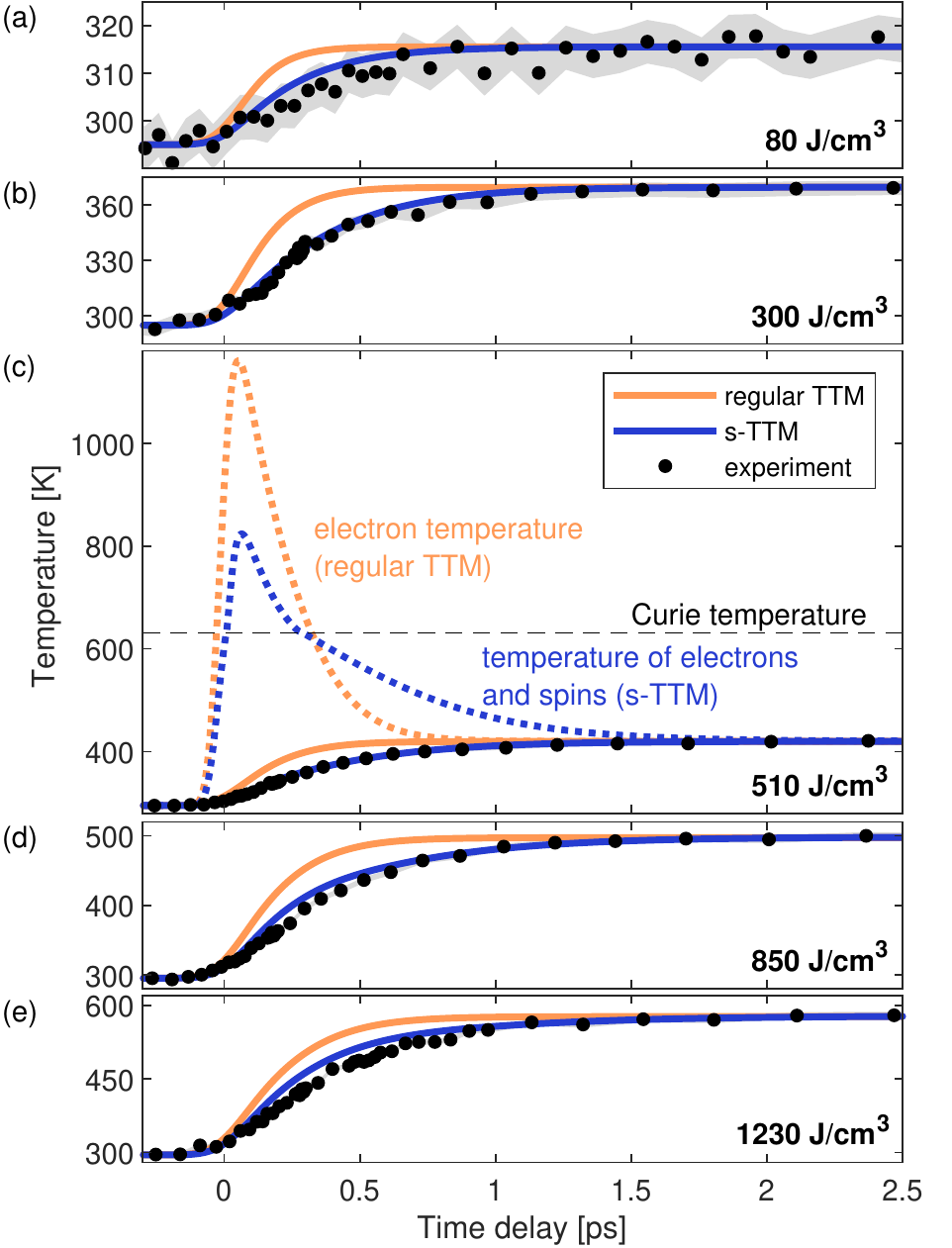}
\caption{Comparison of the experimental results with the regular two-temperature model (TTM) and the modified two-temperature model with infinitely strong electron-spin coupling (\mbox{s-TTM}). The lattice temperature predicted by the regular TTM (solid orange lines) and the s-TTM (solid blue lines) is displayed together with the experimental data for different energy densities (excitation wavelength: \unit[2300]{nm}). Panel~(c) also shows the evolution of the electronic temperatures for the two models (dashed lines). The grey areas represent the standard errors of the experimental data points. Both the TTM and the s-TTM results for the lattice temperature are convolved with a Gaussian (FWHM: \unit[150]{fs}) to account for the pulse duration of the electron pulse. Note that this is less than the convolution width for the single exponential fits of Fig.~\ref{fig:results2300} because the pump pulse duration of \unit[{\raise.17ex\hbox{$\scriptstyle\sim$}}\hspace{1.5pt}80]{fs} is already included in the TTM and s-TTM. The displayed energy densities correspond to the absorbed energy densities of the s-TTM.}
\label{fig:TTM}
\end{center}
\end{figure}

Figure~\ref{fig:TTM} presents the results for the s-TTM (blue curves) and the regular TTM (orange curves) for a range of fluences alongside experimental results (black dots). 
The regular TTM predicts a lattice response that is faster than the experimental results and is therefore inadequate for describing the dynamics of the system. In contrast, the s-TTM yields remarkable agreement with the experimental results, in particular since the lattice response in this model is determined by ab initio results and literature values, without any fit parameters. Clearly, the s-TTM describes the phonon dynamics much better than the regular TTM. This is an indication that a non-negligible amount of energy flows to the spin-system, in agreement with the results of Ref. \cite{2018Tengdin}. This energy transfer leads to a significantly lower transient electronic temperature compared to the regular TTM (see Fig.~\ref{fig:TTM}\hspace{1pt}(e)), which results in a slower electron-lattice-equilibration (see Equations \ref{eq:TTM1},\ref{eq:TTM2}). Note that in general, non-thermal electron and phonon distributions can also lead to a slow-down of the electron-lattice-equilibration. We found that for nickel, non-thermal distributions cannot explain our observations (see Appendix~B for details).

In conclusion, the s-TTM is able to capture the main features of the energy flow to and from magnetic degrees of freedom. It therefore provides a good description of the lattice response. However, a shortcoming of the s-TTM is that it implies quasi-instantaneous demagnetization dynamics, in disagreement with experimental results \cite{2010Koopmans,2018You}. To add a realistic description of the magnetization dynamics, an explicit treatment of the spin system is required, which will be discussed in the next subsection.

\subsection{Atomistic spin dynamics simulations}
\subsubsection{Model and comparison with the experiment} 
In order to consistently describe the evolution of all three subsystems, we employ ASD simulations. These describe the spin system based on a classical Heisenberg model and the stochastic Landau-Lifshitz-Gilbert \mbox{(s-LLG)} equation. The evolution of electron and lattice temperature is based on the TTM with an additional coupling of the spin system to the electron system via the stochastic term of the s-LLG equation. A schematic diagram of the model is displayed in Fig.~\ref{fig:heat_capacities}(e) and further details about the simulations are described in Appendix~C.

Commonly, ASD simulations disregard the energy cost of exciting the spin system since the electron system is considered as a heat bath that acts on the spins. However, in order to account for energy flow between the electron and spin system, the ASD simulations need to be energy-conserving. This was achieved following a similar approach as described in Ref.~\cite{Wienholdt2015}. The energy $\mathcal{H}\{\mathbf{S}_i(t)\}$ of the spin system was monitored during each time step $\Delta t$ of the ASD simulation and the spin energy change $\Delta E_\mathrm{s}$ was calculated:
\begin{equation}
\Delta E_\mathrm{s} = \frac{1}{3}(\mathcal{H}\{\mathbf{S}_i(t + \Delta t)\} - \mathcal{H}\{\mathbf{S}_i(t)\}).
\label{eq:spin_energy}
\end{equation}
Here, $\mathbf{S}_i$ are the individual spins of the simulation and the factor $\frac{1}{3}$ is a correction factor that accounts for the quantized nature of the spins (see Appendix~C for details). The energy change $\Delta E_\mathrm{s}$ of the spin system was subtracted from the electron system, thus coupling the two systems in an energy-conserving way. We note that in our model direct spin-phonon coupling is not considered, which is a reasonable approximation due to the fast time scales of the demagnetization dynamics \cite{2010Koopmans,2018You} and the low magnetocrystalline anisotropy of nickel \cite{MaterialsHandbook}. We therefore modify the TTM equation describing the evolution of the electronic temperature in the following way:

\beq
c_\mathrm{e}\hspace{1pt}\frac{\Delta T_\mathrm{e}}{\Delta t}=G_\mathrm{ep}\hspace{1pt}(T_\mathrm{l}-T_\mathrm{e})+S(t)-\frac{\Delta E_\mathrm{s}}{\Delta t}.
\label{eq:TTM_ASD}
\eeq

\begin{figure}[p!]
\includegraphics[width=1.0\columnwidth]{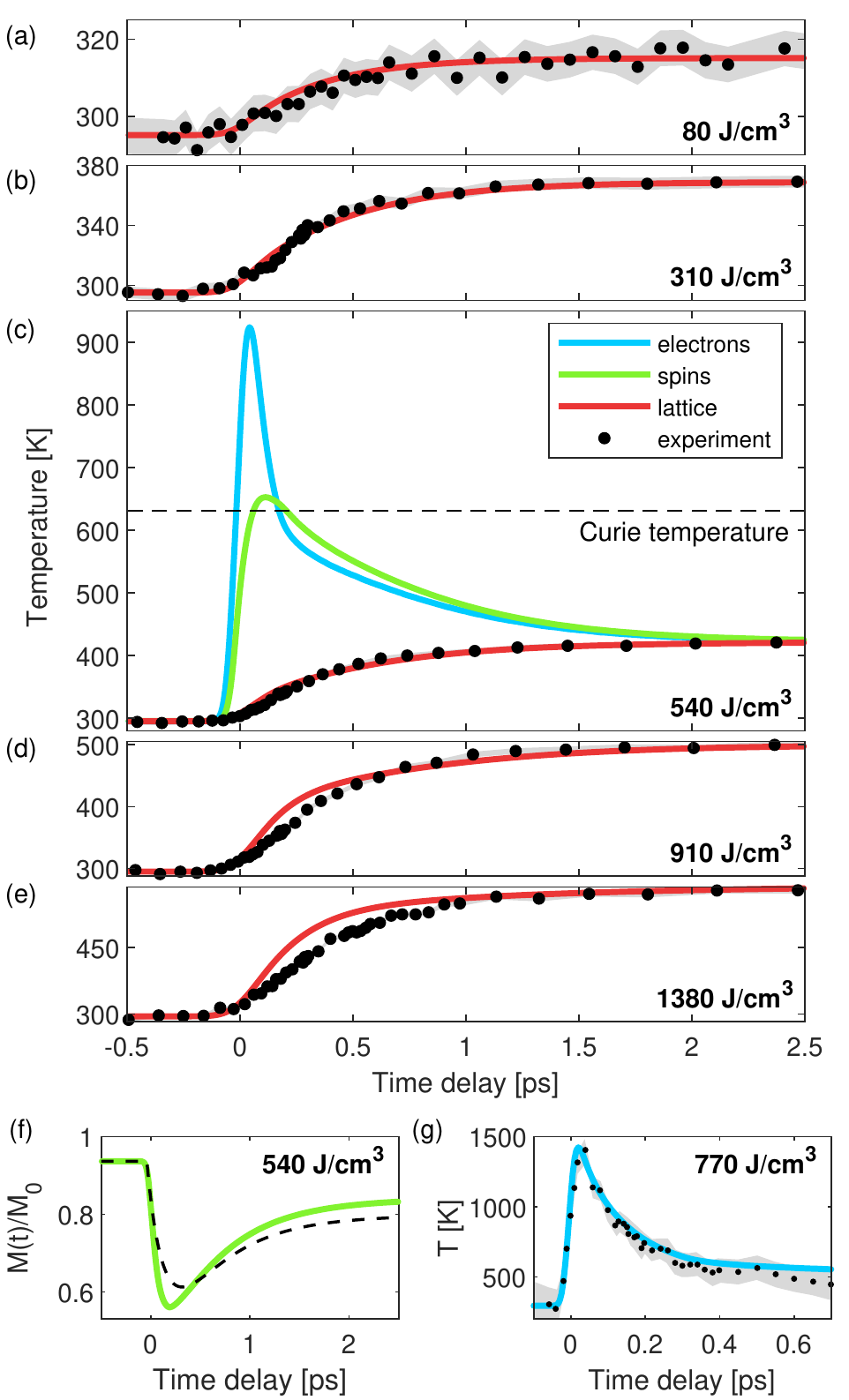}
\caption{Atomistic spin dynamics (ASD) simulations and comparison with the experiment. Panels~(a)-(e) show the comparison between ASD simulations (solid red lines) and experiments (black dots) for different absorbed energy densities. The energy densities are slightly different compared to Fig.~\ref{fig:TTM} due to the different spin heat capacity in the ASD simulations. In the simulations, the pump pulse has a FWHM of \unit[80]{fs}. The results for the lattice temperature are convolved with a Gaussian (FWHM: \unit[150]{fs}) to account for the pulse duration of the electron pulse. The electron-lattice interaction in the simulations is described based on spin-resolved DFT calculations, without fit parameters. Panel~(c) additionally displays the evolution of the electronic (solid blue line) and the spin temperature (solid green line). Panel ~(f) displays the magnetization dynamics predicted by the ASD simulations (solid green line), normalized to the magnetization at $T_\mathrm{s}=\unit[0]{K}$, as well as experimental results from Ref. \cite{2018You} for the same absorbed energy density (dashed black line). Panel~(g) compares the evolution of the electronic temperature in the ASD simulations (solid blue line) to experimental data from Ref. \cite{2018Tengdin} (black dots). In this case, we assumed a shorter pump pulse duration in the simulations (FWHM: \unit[30]{fs}). Note that the sample geometry (film thickness and substrate) was different in the measurements from Refs. \cite{2018You,2018Tengdin}.
The grey shaded areas of Panels~(a)-(e) represent the standard errors of the
data points. The grey shaded area of Panel~(g) represents the errors of the experimental data points from Ref. \cite{2018Tengdin}. }
\label{fig5:ASD_exp}
\end{figure}

Figure~\ref{fig5:ASD_exp}(a)-(e) compares the results of the ASD simulations (solid red lines) using this approach with our experiments (black dots). Similar to the s-TTM, the ASD simulations maintain the excellent agreement with the experimentally measured lattice dynamics, confirming the strong influence of the magnetization dynamics on the lattice dynamics. Note that also in this model, the electron-phonon coupling is not a fit parameter but stems from the spin-resolved DFT calculations.

The main advantage of the ASD simulations is the improved description of the spin system and its magnetization dynamics compared to the s-TTM. This is shown in Fig.~\ref{fig5:ASD_exp}(f), which compares the magnetization dynamics from the ASD simulations with experimental results from Ref. \cite{2018You}. Also for the other fluences, a much better description of the magnetization dynamics is obtained, as shown in Figure~\ref{fig:nonthermal_spins}(a). In addition to the magnetization dynamics, the ASD simulations also yield good agreement with previously reported time- and angle-resolved photoemission (tr-ARPES) measurements of the electronic temperature \cite{2018Tengdin}, shown in Fig.~\ref{fig5:ASD_exp}(g). 

Regarding the lattice dynamics, we find that for very high fluences (Fig.~\ref{fig5:ASD_exp}~(d) and (e)), the agreement of the ASD simulations with the experiments is less good. This can be due to pump-induced changes of the electronic band structure, which are not included in the model and become more pronounced at higher fluences. The results from DFT calculations describe the ground state properties. Hence, the thus obtained electronic band structure and the electron-phonon coupling best describe the weakly perturbed system as produced by low excitation fluences. In addition, the ASD simulations overestimate the spin heat capacity, in particular for high spin temperatures. This leads to an overestimation of the initial energy flow to the spins during demagnetization as well as the energy flow back from the spin system, especially for high fluences. In comparison to the s-TTM, the ASD simulations reach lower quantitative agreement with the high-fluence lattice dynamics. However, the {\sl overall} agreement with the dynamics of all subsystems is significantly improved for all fluences.

For low and moderate absorbed energy densities from \unit[80]{$\frac{\mathrm{J}}{\mathrm{cm}^3}$} to \unit[540]{$\frac{\mathrm{J}}{\mathrm{cm}^3}$}, the ASD simulations yield excellent agreement with the lattice response. The comparison with the electronic, magnetic and lattice responses shows that beyond describing the lattice dynamics, the ASD simulations offer a consistent description of the dynamics of all three subsystems.

\begin{figure}[t!]
\includegraphics[width=1.0\columnwidth]{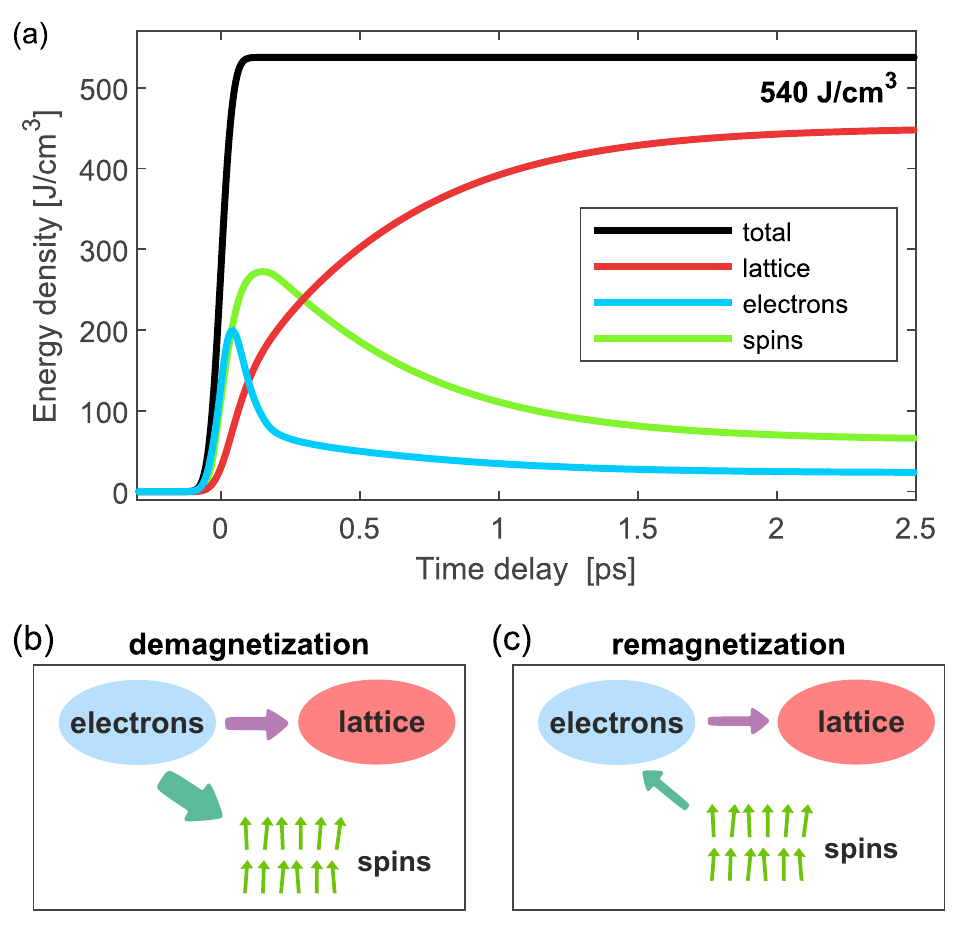}
\caption{Microscopic energy flow between electronic, magnetic and lattice degrees of freedom according to the atomistic spin dynamics (ASD) simulations. Panel~(a) shows how the additional energy after laser excitation is distributed between the three different subsystems as a function of time. The black line corresponds to the total additional energy in the material, demonstrating that energy is conserved in the model. Panel~(b) visualizes the energy flow during the demagnetization. There is a large energy flow from the electrons to the spin system and as well as energy flow from electrons to the lattice. Panel~(c) shows the energy flow during remagnetization. Energy flows back from the spins to the electrons. In addition, energy flows from the electrons to the lattice, such that the electron as well as the spin energy decreases while the lattice energy increases.}
\label{fig:ASD_energy_flow}
\end{figure}
\subsubsection{Energy flow dynamics}
To highlight and discuss some of the key advantages of the ASD simulations and to gain further insights into the energy flow between the different subsystems, we now discuss the details of the temperature and energy dynamics. For this discussion, we also calculate a spin temperature (see Appendix~C for details). Note that the spin system is not always in internal thermal equilibrium during the simulations, as will be discussed later. Figure~\ref{fig5:ASD_exp}(c) displays the temperature dynamics of electrons (blue), phonons (red) and spins (green) after the initial laser excitation for an absorbed energy density of \unit[540]{$\frac{\mathrm{J}}{\mathrm{cm}^3}$}. The electron temperature increases rapidly when the laser pulse excites the sample. Contrary to the assumptions made for s-TTM, however, the spin temperature does not follow the electron temperature instantly. Instead, the spin temperature increase is slower and delayed due to the finite coupling between electrons and spins. After \unit[{\raise.17ex\hbox{$\scriptstyle\sim$}}\hspace{1.5pt}160]{fs}, the two subsystems have reached a similar temperature, and they cool down at similar rates while the lattice still heats up. Finally, thermal equilibrium is reached after {\raise.17ex\hbox{$\scriptstyle\sim$}}\hspace{1.5pt}\unit[2-2.5]{ps}. 

In addition to the temperatures, the ASD simulations also provide the energy dynamics of the different subsystems, shown in Fig.~\ref{fig:ASD_energy_flow}(a). After the initial laser excitation, the total additional energy in the system (solid black line) stays constant and energy is only transferred between subsystems. Initially, the electron system (solid blue line) absorbs all of the deposited energy. The rise of the electronic temperature initiates the demagnetization dynamics and energy immediately starts flowing to the spin system (solid green line). Here, we identify the key feature that is not captured by the regular TTM: Already shortly after excitation, the spin system contains more energy than the electron system, which leads to the significant slow-down of the lattice dynamics. 

The energy flow during demagnetization is schematically depicted in Fig.~\ref{fig:ASD_energy_flow}(b). In addition to the energy flow to the spin system, energy also flows to the lattice (solid red line), although at a lower rate. After \unit[{\raise.17ex\hbox{$\scriptstyle\sim$}}\hspace{1.5pt}150]{fs}, the energy flow to the spin system stops due to the lower electronic temperature. This initiates the remagnetization dynamics. Energy starts flowing back from the spin system to the electrons, which is visualized in Fig.~\ref{fig:ASD_energy_flow}(c). Energy also flows from the electrons to the lattice, such that in total, the electrons lose further energy, although at a much slower rate than during the demagnetization (see Fig.~\ref{fig:ASD_energy_flow}(a)). Note that there is no direct energy flow from the spins to the lattice in the model, but the net energy flow from spins to the lattice is indirect via the electrons. These processes continue until thermal equilibrium is established.

Note that the experiments on the three different subsystems in Fig.~\ref{fig5:ASD_exp} were performed under different experimental conditions. Therefore, the deviations of the experimental data from the simulations cannot be directly interpreted in terms of energy flow.
\\
\\
\subsubsection{Non-thermal spin dynamics}
Next, in order to gain further insights into the non-equilibrium behavior of the spin system, we analyze the ASD simulation results for the spin system in detail. The simulations provide the spin temperature, the spin energy as well as the magnetization simultaneously. By comparing these three quantities, further conclusions on the non-equilibrium spin system can be drawn. First of all, note that despite the fact that the spin temperature in Fig.~\ref{fig5:ASD_exp}(c) rises above the Curie temperature, the system does not demagnetize completely, as displayed in Fig.~\ref{fig5:ASD_exp}(f). This demonstrates that on short time scales after laser excitation, the spin system is not in internal thermal equilibrium. 

\begin{figure}[bth]
\includegraphics[width=1.0\columnwidth]{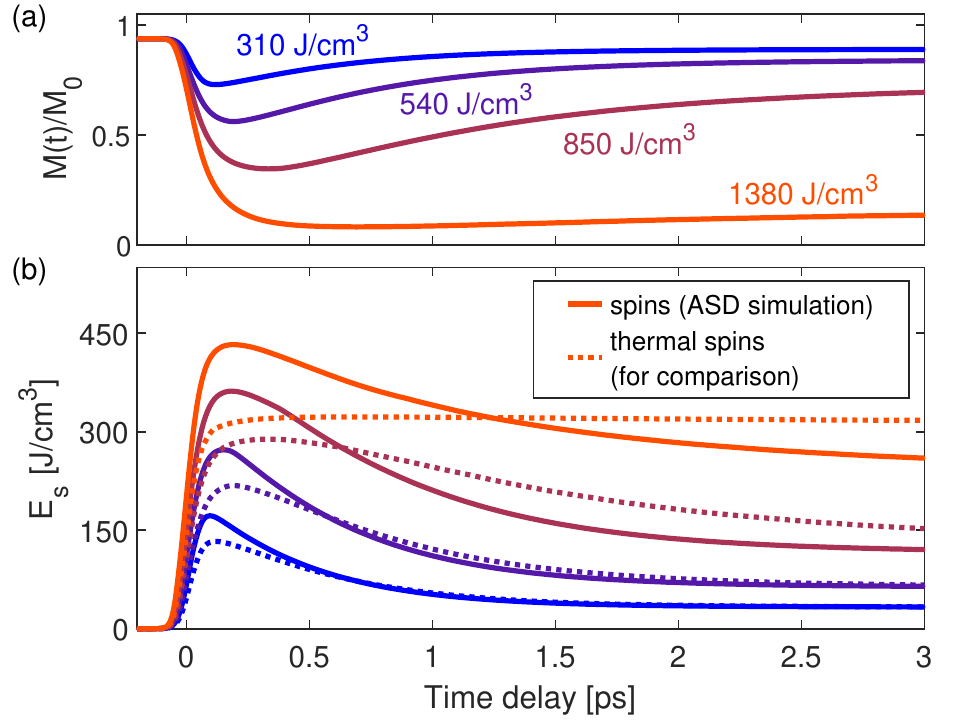}
\caption{Non-equilibrium spin dynamics. (a) Magnetization dynamics from the ASD simulations for several excitation densities of our experiments. (b) Energy content of the spin system as a function of pump-probe delay for several excitation densities (solid lines). For comparison, the dashed lines show the energy content of a hypothetical, thermalized spin system with the magnetization dynamics from the ASD simulations (shown in Panel (a)).}
\label{fig:nonthermal_spins}
\end{figure}

To understand the characteristics of this transient non-thermal state, we analyze the magnetization and energy content of the spin system. The magnetization dynamics following laser excitation are displayed in Fig.~\ref{fig:nonthermal_spins}(a) for several excitation densities. The corresponding additional spin energy content is shown in Fig.~\ref{fig:nonthermal_spins}(b) (solid lines).
We compare the evolution of these two quantities after laser excitation to the case in which the spin system is heated quasi-statically. The latter case is obtained from the ASD simulations by increasing the energy of the system in small steps and waiting for the system to reach equilibrium after each step (see Appendix~C for details on heat capacities and statistics). By comparing the simulations of the laser-excited dynamics to the equilibrium relationships, we find that on short time scales after laser excitation, the ASD simulations predict a spin energy content that is higher than in equilibrium for the same magnetization. This is visualized by the dashed lines in Fig.~\ref{fig:nonthermal_spins}(b), which represent the energy content of a hypothetical, thermalized spin system undergoing the magnetization dynamics predicted by the ASD simulations (shown in Fig.~\ref{fig:nonthermal_spins}(a)). The comparison with the actual spin energy indicates that on short timescales, the spin system is in a transient non-thermal state with a large amount of high-energy spin excitations, in agreement with previous experimental results \cite{2018Tengdin}.

This behavior is analogous to non-thermal phonon distributions: In cases in which high-energy phonons couple strongly to the lattice, the atomic displacements can be relatively small compared to the lattice energy content on short time scales \cite{2016Waldecker}, because the equilibrium relationship between atomic displacements and lattice energy content is not applicable. Similarly, if the distribution of spin excitations differs from thermal equilibrium, the equilibrium relationship between magnetization and energy content of the spin system is not applicable. In the ASD simulations, the energy transfer from electrons to the spin system creates mostly high-energy spin excitations due to the localized nature of the electron-spin interaction. During the thermalization of the spin system, these excitations then decay into more delocalized spin waves with a larger magnetization reduction per energy. The lifetime of a spin wave mode can be estimated by $\tau \approx \frac{1}{2\alpha\omega }$ \cite{Gurevich}, where $\omega$ is the angular frequency of the spin wave and $\alpha$ is the Gilbert damping. In nickel, for the high-energy spin waves at the Brillouin zone boundary \cite{Niesert}, this corresponds to a lifetime of \unit[{\raise.17ex\hbox{$\scriptstyle\sim$}}\hspace{1.5pt}70]{fs}.
Consequently, the relationship between magnetization and spin energy relaxes towards the thermal relationship within a few hundred femtoseconds. 

On longer time scales, the behavior of the magnetization reverses: the magnetization recovery is delayed compared to the energy flow out of the spin system, particularly for high fluences. We find that if the magnetization is strongly reduced, the spin-system remains non-thermal for several picoseconds. This behavior is in agreement with previous ASD simulation results and was attributed to domain formation \cite{Kazantseva2007}.

The comparison of the laser-induced dynamics to quasi-static heating highlights a main advantage of the ASD simulations: in contrast to temperature models, non-thermal states of the spin system can also be described, since the evolution of the spins is simulated directly. The non-equilibrium behavior of the spin system predicted by the ASD simulations results in good agreement of the model with the experimentally measured lattice dynamics as well as the magnetization dynamics.

Consequently, using ASD simulations we have improved the theoretical description in two key aspects compared to the s-TTM: First, the magnetization dynamics are described realistically since we no longer assume infinite electron-spin coupling, which leads to instantaneous demagnetization. Second, we no longer use the equilibrium spin heat capacity to describe the spin system in this highly non-equilibrium scenario. Instead, we directly calculate the energy content of the spin system in the ASD simulations. These improvements allow for an excellent quantitative description of the experimentally measured lattice dynamics and provide a consistent model for the dynamics of the three subsystems after laser excitation.

Unlike many previous demagnetization models, our approach has the advantage that the parameters for the ASD simulations stem either from ab-initio DFT calculations or are directly linked to measurable quantities, such as the Curie temperature. The avoidance of fit parameters, in combination with the comparison of the model to measurements of all subsystems, is the key to a consistent description of the laser-induced dynamics.

The sole parameter that is only indirectly accessible through experiments is the Gilbert damping parameter $\alpha$. Here, we use $\alpha$=0.01, which yields good agreement with the lattice dynamics and is consistent with literature \cite{damping1,damping2}. We tested different values for $\alpha$ from 0.005 to 0.02, shown in Appendix~C, and found good overall agreement to the experimental data, therefore showing the robustness of the model regarding variations of $\alpha$. Since experimental results can always be influenced by transport or sample-specific effects, a more precise result for $\alpha$ could be obtained by measuring the dynamics of several subsystems on the same sample, ideally on a freestanding thin film. Furthermore, since $\alpha$ is a phenomenological constant that comprises several microscopic effects, additional accuracy could be gained by disentangling these microscopic effects in a future model.
\\

\section{Summary and conclusions}

In this work, we combined direct experimental measurements of the lattice response with first-principles calculations of the electron-phonon interaction and atomistic spin dynamics (ASD) modelling in order to obtain a full picture of the dynamics in ferromagnetic nickel following laser excitation. The combination of theory and experiment enabled us to study the influence of the energy cost of demagnetization on the lattice dynamics. We found that energy flow to and from the spin system leads to a significant slow-down of the lattice dynamics. The spin system is the dominant heat sink in the initial few hundreds of femtoseconds. Consequently, it is paramount to include the energy flow to and from the spin system in any description of the laser-induced dynamics. 

In case only the lattice dynamics are of interest, a modified TTM employing electron-phonon coupling from first-principles calculations and incorporating infinitely strong electron-spin coupling (s-TTM) suffices. The agreement of the s-TTM with the measured lattice dynamics proved to be vastly superior to  that of the regular TTM. 

A consistent description of the coupled energy flow between all three subsystems and of the magnetization dynamics is obtained with energy-conserving ASD simulations. Like the s-TTM, the ASD simulations are based on first-principles calculations, thus minimizing the use of fit parameters. The comparison with available experimental data for the electronic, lattice and spin dynamics shows that the ASD simulations achieve a quantitative description of all three subsystems. In the future, the precision of this comparison could be improved further by measuring the response of all three subsystems on identical samples. 

Both the s-TTM and the ASD simulations unambiguously demonstrate the strong influence of the magnetization dynamics on the lattice dynamics, highlighting the importance of considering their coupling in a full description of the material's response to laser excitation.

In addition, the ASD simulations predict that shortly after excitation, the spin system is in a transient non-thermal state and absorbs more energy compared to thermal equilibrium. This finding is corroborated by the excellent agreement of the ASD simulations to the lattice, the electron and the magnetization dynamics. Therefore, our findings indicate that in order to describe both the microscopic energy flow and the magnetization dynamics accurately, an approach that considers non-thermal spin dynamics is necessary.

We expect our findings to be valid for other magnetic metals as well, in particular for other itinerant 3{\sl d} ferromagnets, but also for antiferromagnetic or ferrimagnetic metals. Furthermore, a quantitative description of the microscopic energy flow in ferromagnetic metals is valuable for the design of high-speed spintronic structures, since the functionality of magnetic heterostructures depends on their behavior in non-equilibrium states. This, in turn, is governed by the microscopic energy flow and magnetization dynamics within each component as well as interfacial coupling.

\FloatBarrier

\section*{Acknowledgement}
This work was funded by the Deutsche Forschungsgemeinschaft (DFG) through SFB/TRR 227  "Ultrafast Spin Dynamics" (Projects B07, A08, and A02) and through the Emmy Noether program under Grant No. RE 3977/1, by the European Research Council (ERC) under the European Union’s Horizon 2020 research and innovation program (Grant Agreement Number ERC-2015-CoG-682843), and by the Max Planck Society. H.S.~acknowledges support by the Swiss National Science Foundation under Grant No.~P2SKP2\textunderscore184100. Y.Q.~acknowledges support by the Sino-German (CSC-DAAD) Postdoc Scholarship Program (Grant No.~57343410).

\section*{Appendix~A: DFT calculations}
\begin{figure}[bth]
\begin{center}
\includegraphics[width=1.0\columnwidth]{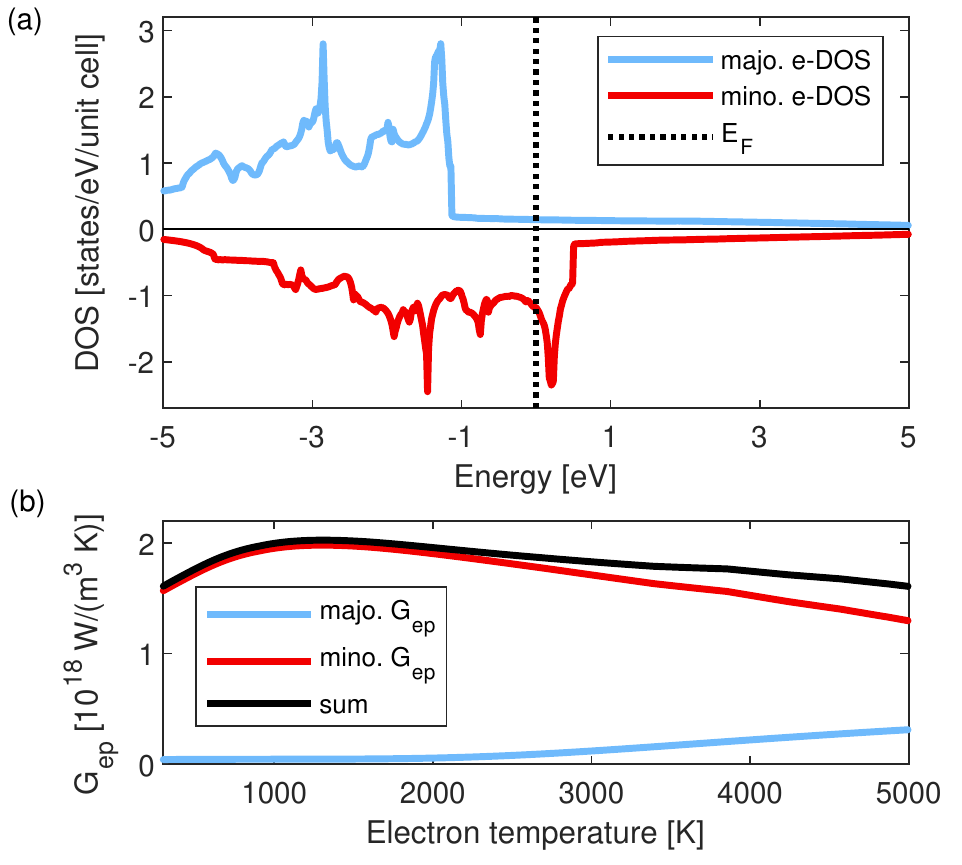}
\caption{Spin-polarized electronic density of states (DOS) and electron phonon coupling of nickel, calculated using spin-resolved DFT. (a) Electronic DOS. The Fermi level is marked with a dashed black line. (b) Electron-phonon-coupling parameter $G_\mathrm{ep}$ as a function of electron temperature. The majority $G_\mathrm{ep}$ (blue), the minority $G_\mathrm{ep}$ (red) and their sum (black) is displayed.}
\label{fig:s1-DFT}
\end{center}
\end{figure}

The calculations of the electron-phonon energy transfer rates were performed using the DFT code ABINIT \cite{Gonze1997,Gonze1997a,Gonze2009,Gonze2016,Bottin2008}. The norm-conserving electron-ion pseudopotential was generated using the FHI package \cite{FUCHS199967} and is of GGA-PBE type \cite{Perdew:1996}. 10 electrons were treated explicitly and 18 electrons were frozen in the core. The plane wave expansion of the electronic wavefunction had a cutoff of $50$ Ha. 20 electronic bands were calculated. These bands are calculated with Fermi occupation featuring a smearing of $0.001$ Ha. An unshifted k-point grid of $32\times 32\times 32$ points was used. The experimental lattice constant of the fcc lattice of $d=6.6594$ a$_B$ (=\unit[3.5240]{\AA}) was used. Figure~\ref{fig:s1-DFT}(a) shows the result for the spin-polarized electronic DOS. The DFT calculation predicts a magnetic moment of \unit[0.815]{$\mu_\mathrm{B}$}, which is larger than the experimentally measured value of \unit[0.616]{$\mu_\mathrm{B}$} \cite{1968Danan}. This overestimation mainly affects the minority DOS at the Fermi level. We therefore tested its effects on our models by shifting the minority DOS to lower energies in several steps, until the maximum of the minority DOS coincides with the Fermi level. We then calculated TTM results based on these shifted DOS. Since the differences in the lattice responses are small, we conclude that the overestimation of the magnetic moment has no significant effect on our results. Regarding the phonons, the shape and energy range of the calculated phonon DOS (not shown) agree well with neutron scattering experiments \cite{2007Kresch}. The phonon DOS is used to calculate the lattice heat capacity, resulting in excellent agreement with experimental results \cite{1981Meschter}.

To obtain the electron-phonon coupling $G_\mathrm{ep}$, the spin-resolved electron-phonon matrix elements were computed as described in Ref. \cite{Verstraete} for a $8\times 8\times 8$ grid of q-points. From the results, we extracted the Eliashberg functions (also phonon-branch resolved) for majority and minority electrons. The electron-phonon couplings as well as the electronic heat capacities were then calculated as in Ref. \cite{2016Waldecker}. The result for the electron-phonon coupling is displayed in Fig.~\ref{fig:s1-DFT}(b). In the calculation of the spin-resolved electron-phonon coupling and electronic heat capacities, we assume that the particle number is conserved within each spin type. In practice, for the electron temperatures reached in our experiments, the chemical potential shifts are small, and thus the differences between assuming two separate chemical potentials or assuming a common chemical potential are small. For the temperature models as well as the ASD simulations, we use the sum of majority and minority $G_\mathrm{ep}$ (black curve of Figure~\ref{fig:s1-DFT}(b)). Correspondingly, the electronic heat capacity used in the models is also the sum of minority and majority electronic heat capacity. The results of the DFT calculation are available on a data repository \mbox{\cite{rodare_data}}.

We note that our result for the electron-phonon coupling is significantly larger compared to results by Lin et al. \cite{2008Lin}, but similar to a spin-resolved calculation by Ritzmann et al. \cite{2019Ritzmann}. We also find significant differences compared to the values used in existing demagnetization models: In the original 3TM by Beaurepaire et al. \cite{1996Beaurepaire}, a much smaller value of \unit[$8\cdot10^{17}$]{W/(m$^3$K)} is used, resulting in a slower lattice response compared to our experiments. In the M3TM \cite{2010Koopmans}, the value for $G_\mathrm{ep}$ is \unit[$4.05\cdot10^{18}$]{W/(m$^3$K)}, which differs from our result by more than a factor of 2. In addition, the heat capacities are different. In the $\mu$T model \cite{2014Mueller}, the same $G_\mathrm{ep}$ of \mbox{\unit[$1\cdot10^{18}$]{W/(m$^3$K)}} is used for majority and minority carriers, whereas the $G_\mathrm{ep}$ from our ab initio calculations shows significant differences between majority and minority carriers. 

\section*{Appendix~B: The influence of non-thermal electron and phonon distributions}

\begin{figure}[bth]
\begin{center}
\includegraphics[width=1.0\columnwidth]{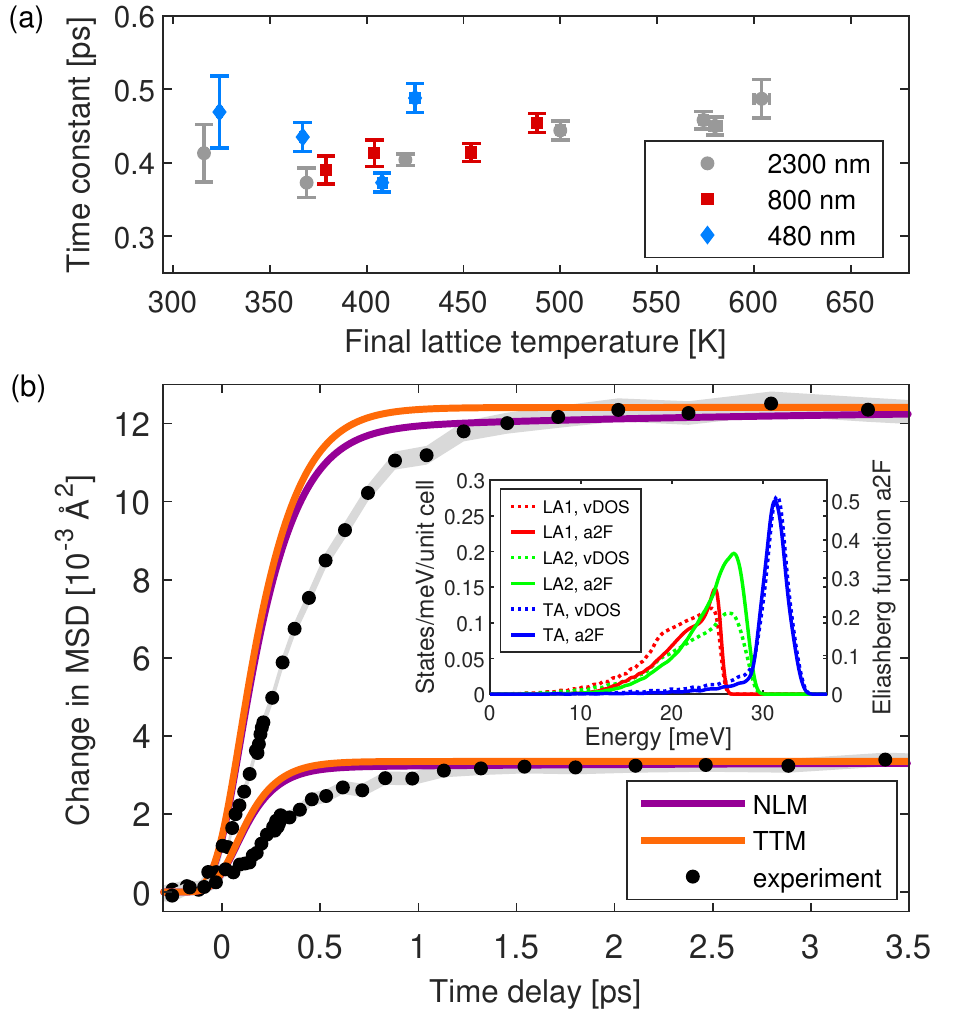}
\caption{Experimental and theoretical results regarding electron and phonon thermalization. (a) Time constants of electron-lattice equilibration for different excitation wavelengths, obtained by single-exponential fits of the experimental data. The grey dots are the same data as in the inset of ~\ref{fig:results2300}, shown again for comparison. The error bars represent the standard errors from the single exponential fits. (b) Comparison of two-temperature model (TTM) results with non-thermal lattice model (NLM) results for two different fluences. Experimental data for \unit[2300]{nm} excitation wavelength are also shown. The grey shaded areas represent the errors of the experimental data. The inset shows the Eliashberg function (solid lines, sum of majority and minority Eliashberg function) and the phonon DOS (dashed lines) projected onto the three phonon branches.}
\label{fig:s1-nonthermal}
\end{center}
\end{figure}

The TTM relies on the assumption that electrons and phonons are each in a thermal state, which is not necessary fulfilled shortly after laser excitation. For electrons, in metals, thermalization is typically rather efficient due to the large phase space for electron-electron scattering. In the case of nickel, there is experimental evidence for efficient electron-electron scattering \cite{2018Tengdin}. In addition, to test if our measured lattice dynamics are influenced by non-thermal electrons, besides the experiments with \unit[2300]{nm}, we also performed experiments with \unit[800]{nm} and \unit[480]{nm} excitation wavelength and compared the lattice dynamics. Figure~\ref{fig:s1-nonthermal}(a) shows the time constants of a single exponential fit to the lattice temperature for these three wavelengths and different excitation densities. No dependence of the lattice dynamics on the wavelength is observed. From this, we conclude that electrons thermalize on timescales significantly faster than the timescales of electron-phonon equilibration. Otherwise, we would expect an influence of the photon energy on the lattice dynamics, since different initial states are excited and different electronic states have different lifetimes for electron-phonon scattering. Hence, we conclude that it is justified to assume a thermalized electron distribution in our models.

On the other hand, for phonons, the assumption of a thermalized distribution is often more problematic \cite{2019Ritzmann,2016Waldecker}. We investigated the influence of non-thermal phonon distributions on our observable, the MSD, using a non-linear lattice model (NLM) \cite{2016Waldecker}. 
The three different phonon branches are treated as individual subsystems in order to account for energy redistribution between them. For this, we calculate the branch-projected phonon DOS and Eliashberg functions, shown in the inset of ~\ref{fig:s1-nonthermal}(b). We don't take into account direct phonon-phonon coupling, which means that the equilibration between the phonon branches is mediated by electron-phonon coupling only. The comparison between TTM and NLM is displayed in Fig.~\ref{fig:s1-nonthermal}(b) and shows only small differences between the lattice temperatures predicted by the two models. In addition, a previously reported model predicts only minor deviations of the electronic temperature evolution compared to a TTM for nickel \cite{2019Ritzmann}. There are experimental observations of phonon thermalization processes in nickel \cite{2019Maldonado}, mostly observed in the range of \unit[1-4]{ps} after laser excitation. Since we don't observe any significant MSD changes during this period (see Figure~\ref{fig:results2300}), we conclude that the effect of these phonon thermalization processes on the MSD is small, and that the sub-picosecond dynamics that we observe correspond to electron-phonon equilibration. Based on these theoretical and experimental results, we conclude that in the case of nickel and for the purpose of describing energy flow between electrons and the lattice, a thermalized phonon population is a reasonable approximation.

\section*{Appendix~C: Atomistic spin dynamics simulations}

\begin{figure}[t!]
\includegraphics[width=1\columnwidth]{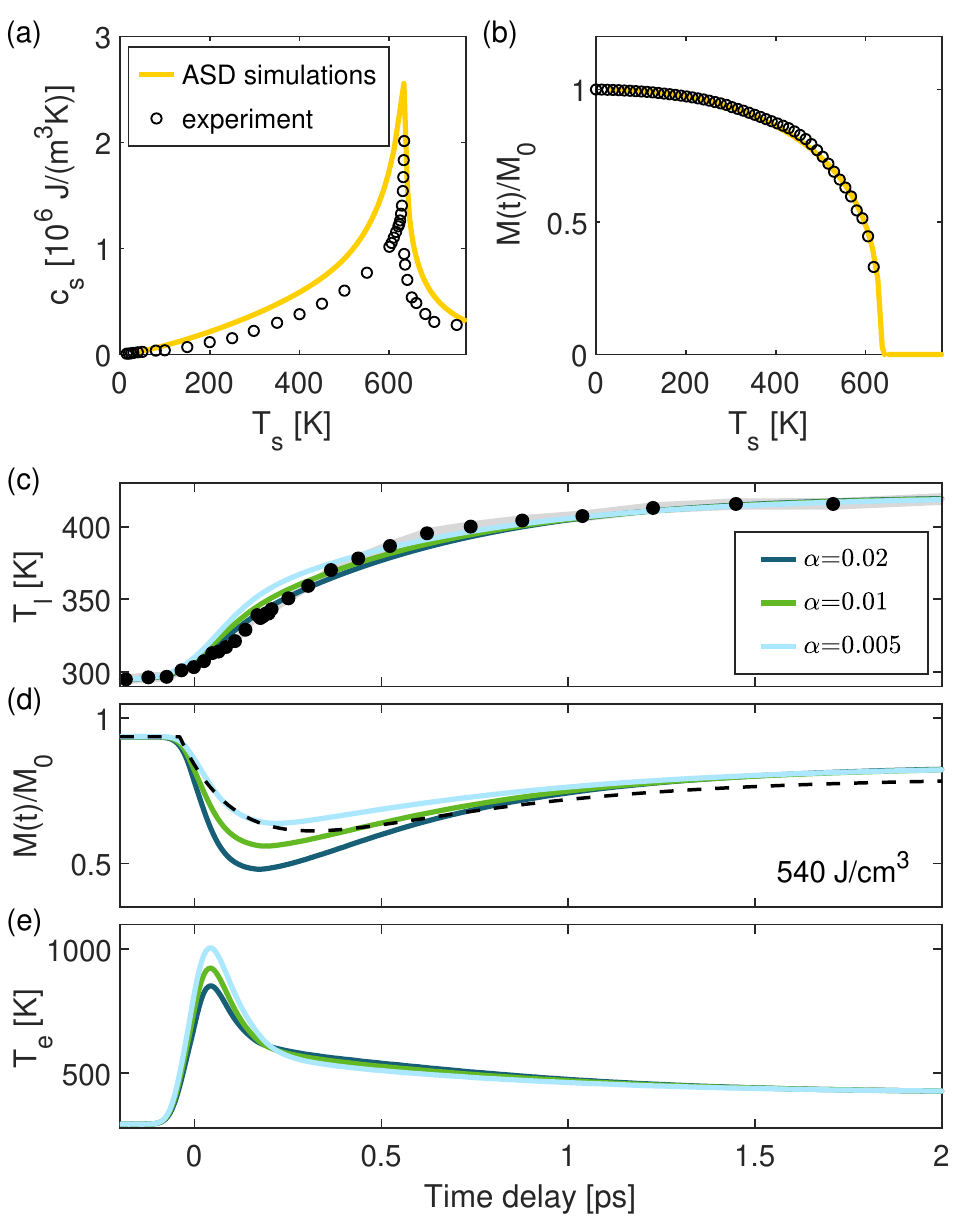}
\caption{Atomistic spin dynamics (ASD) simulation results for equilibrium and non-equilibrium conditions. (a) Comparison between experimentally measured equilibrium heat capacity (black circles) and the simulated equilibrium heat capacity (yellow line). The experimentally measured spin heat capacity corresponds to the heat capacity of electrons and spins \cite{1981Meschter} minus the electronic heat capacity from our DFT calculations. (b) Comparison between the experimentally measured magnetization curve as a function of temperature from Ref. \cite{Crangle1971} (black circles) to the simulation (yellow line). (c) Experimentally measured lattice dynamics (black dots) and ASD simulation results (solid lines) for different values of the Gilbert damping parameter $\alpha$. The absorbed energy density is \unit[540]{$\frac{\mathrm{J}}{\mathrm{cm}^3}$}. The grey shaded area represents the errors of the experimental data. (d) Magnetization dynamics predicted by the ASD simulations for different values of $\alpha$ (solid lines). The dashed black line corresponds to the experimental magnetization dynamics for the same absorbed energy density of \unit[540]{$\frac{\mathrm{J}}{\mathrm{cm}^3}$} from Ref. \cite{2018You}. (e) Evolution of the electronic temperature for different values of $\alpha$ according to the ASD simulations. Note that in addition to $\alpha$, the initial rise of the electronic temperature also depends on the pump pulse duration (here: \unit[80]{fs} FWHM).}
\label{fig:s3_ASD_equilibrium}
\end{figure}

In the ASD simulations, the spin system is described using a classical Heisenberg Hamiltonian:
 \begin{equation}   
\mathcal{H}= - \sum_{i < j} J_{ij} \hspace{1pt}\mathbf{S}_i \cdot \mathbf{S}_j - \sum_{i} d_z S_z^2.
\label{eq:Ham}
\end{equation}
Here $\mathbf{S}_i $ represents a unit vector describing the direction of the local magnetic moment at site $i$.
Each spin $\mathbf{S}_i$, couples to its neighboring spins $\mathbf{S}_j$ via the exchange constant $J_{ij}= 2.986\cdot 10^{-21}$~J. We use a simple cubic lattice structure with a spin volume of $V_\text{s} = 10.94$ $\mathring{A}^3$. We tested different lattice structures and found that this has no significant effect on our results. To obtain the correct spin energy from Equation~\ref{eq:Ham}, a correction factor of 1/3 is necessary (see  Equation~\ref{eq:spin_energy}). This accounts for the fact that the spins are quantized in reality (s $\approx$ 1/2 for nickel), but described with the classical Heisenberg Hamiltonian (s=$\infty$). The relationship between the exchange constant $J_{ij}$ and the Curie temperature $T_\mathrm{c}$ depends on the quantum number $s$. For a simple cubic system with only nearest neighbor interaction \cite{White},
\begin{equation}
    J=\frac{s^2}{s(s+1)}\cdot\frac{3k_\mathrm{B}}{T_\mathrm{c}}.
\end{equation}
Consequently, to obtain a good description of both the Curie temperature and the energy content of a spin system with finite $s$, a factor of $\frac{s^2}{s(s+1)}$ ($\frac{1}{3}$ for $s=\frac{1}{2}$) needs to be considered. The second term of Equation~\ref{eq:Ham} describes the on-site anisotropy with easy-axis along the $z$ axis and a constant anisotropy energy, \unit[$d_z=5\cdot10^{-24}$]{J}. 
The ASD-simulations are performed by solving the stochastic-Landau-Lifshitz-Gilbert equation (s-LLG) numerically using the Nvidia CUDA C-API \cite{CudaPrograming,Nowak2007}.
\begin{equation}
\frac{(1+\alpha^2)\mu_{s}}{\gamma}\frac{\partial \mathbf{S}_i}{\partial t} = - \left( \mathbf{S}_i \times \mathbf{H}_i \right) - \alpha  \left( \mathbf{S}_i \times \left(\mathbf{S}_i \times \mathbf{H}_i \right) \right).
\label{eq:llg}
\end{equation}
$\gamma=1.76 \cdot 10^{11}$ is the gyromagnetic ratio and $\mathbf{H}_i$ is the effective field (see below). For the magnetic moment $\mu_{s}$ we use the literature value of \unit[0.616]{$\mu_\text{B}$} \cite{1968Danan}, which contains the spin as well as the (smaller) orbital contribution. The phenomenological Gilbert damping $\alpha$ determines the coupling strength of the spin system to the electron system and thus the energy transfer rate between these two subsystems.
A Langevin thermostat is included, by adding a field-like stochastic term $ \boldsymbol{\zeta}_i$ to the effective field $\mathbf{H}_i= \boldsymbol{\zeta}_i(t) - \frac{\partial \mathcal{H}}{\partial \mathbf{S}_i}$. The added noise term has white noise properties~\cite{Atxitia2009}:
\begin{equation}
\langle \boldsymbol{\zeta}_i(t) \rangle = 0 \quad \text{and} \quad \langle \boldsymbol{\zeta}_i(0) \boldsymbol{\zeta}_j(t) \rangle = 2 \alpha k_\text{B} T_{\rm{el}} \mu_{s} \delta_{ij}\delta(t)/\gamma.
\label{eq:noise-correlator}
\end{equation}
The electron temperature $T_{\rm{el}}$ is therefore used to scale the noise and has a direct impact on the spin dynamics via the stochastic field $\boldsymbol{\zeta}(t)$ entering the s-LLG.
The s-LLG is solved for system sizes of several million spins. These large systems yield minimal boundary effects and provides a large enough number of spins for calculating macroscopic parameters.
While showing excellent qualitative agreement with experiments, due to their classical character ASD simulations are typically unable to quantitatively reproduce thermodynamic properties such as the heat capacity or the temperature-dependent equilibrium magnetization.
To counteract this shortcoming, we make use of a rescaled temperature model \cite{Evans2015}. A modified electron temperature $T_\text{sim}$, based on $T_c$ and a material dependent factor $\beta=2.322$ is used:
\begin{equation}
T_\text{sim}= T_c \left(\frac{T_\text{el}}{T_c} \right)^\beta.
\end{equation}
This correction allows us to reproduce experimentally measured quantities such as the temperature-dependent equilibrium magnetization curve and the heat capacity (see Figure~\ref{fig:s3_ASD_equilibrium}(a) and (b)). For the temperature-dependent equilibrium magnetization (Figure~\ref{fig:s3_ASD_equilibrium}(b)) we obtain excellent agreement with experimental values. The spin heat capacity (Figure~\ref{fig:s3_ASD_equilibrium}(a)) is overestimated due to the classical nature of the spins in the ASD-simulations.

The spin temperature in ASD simulations can be calculated through the instantaneous spin configuration following Ref.~\cite{Ma2010}:
\begin{equation}
T_s=\frac{\mu_\mathrm{s}\left \langle  \sum_i \vert \mathbf{S}_i \times \mathbf{H}_i {\vert}^2 \right \rangle}{2k_{\mathrm{B}} \langle  \sum_i \mathbf{S}_i \cdot \mathbf{H}_i \rangle}.
\label{eq:Teff}
\end{equation}
Here $\mathbf{S}_i$ and  $\mathbf{H}_i$ represent the normalized spin variable and effective field at the lattice site $i$. The spin temperature in Equation~\eqref{eq:Teff} is defined as the ratio between the entropy and energy of spin degrees of freedom,  $\mathbf{S}_i \times \mathbf{H}_i$ and $\mathbf{S}_i \cdot \mathbf{H}_i$, respectively. Note that despite this definition of a spin temperature, the spin system is not always in internal thermal equilibrium during the simulations. The values for the electronic heat capacity, lattice heat capacity and electron-phonon coupling are taken from the DFT calculations described earlier. The laser pulse is assumed to be Gaussian, with a FWHM of \unit[80]{fs} and its peak intensity at $t=0$.

Figure~\ref{fig:s3_ASD_equilibrium}(c)-(d) shows the ASD simulation results for different values of the Gilbert damping parameter $\alpha$. Figure~\ref{fig:s3_ASD_equilibrium}(c) displays the lattice temperature according to the ASD simulations alongside our experimental result while Figure~\ref{fig:s3_ASD_equilibrium}(d) shows the magnetization dynamics from the simulations together with experimental results from Ref. \cite{2018You}. Figure~\ref{fig:s3_ASD_equilibrium}(e) presents the evolution of the electronic temperature according to the ASD simulations.

\FloatBarrier
\pagebreak

\bibliography{main}

\end{document}